\documentclass[%
onecolumn,
10pt,
nofootinbib,
 amsmath,amssymb,
 aps,
 a4paper,
 notitlepage,
prd
]{revtex4-1}

\usepackage{graphicx}
\usepackage{dcolumn}
\usepackage{bm}
\usepackage{amsmath}
\usepackage{color}
\usepackage[utf8]{inputenc} 

\begin{document}

\title{On the viability of bigravity cosmology}

\author{Michael Kenna-Allison}
\email{michael.kenna-allison@port.ac.uk}
\affiliation{Institute of Cosmology and Gravitation, University of Portsmouth\\ Dennis Sciama
Building, Portsmouth PO1 3FX, United Kingdom}

\author{A. Emir G\"umr\"uk\c{c}\"uo\u{g}lu}
\email{emir.gumrukcuoglu@port.ac.uk}
\affiliation{Institute of Cosmology and Gravitation, University of Portsmouth\\ Dennis Sciama
Building, Portsmouth PO1 3FX, United Kingdom}

\author{Kazuya Koyama}
\email{kazuya.koyama@port.ac.uk}
\affiliation{Institute of Cosmology and Gravitation, University of Portsmouth\\ Dennis Sciama
Building, Portsmouth PO1 3FX, United Kingdom}

\date{\today}

\begin{abstract}
We revisit the question of viability of bigravity cosmology as a candidate for dark energy. In the context of the low energy limit model, where matter couples to a single metric, we study linear perturbations around homogeneous and isotropic backgrounds to derive the Poisson's equation for the Newtonian potential. Extending to second order perturbations, we identify the Vainshtein radius below which non-linear scalar self interactions conspire to reproduce GR on local scales. We combine all of these results to determine the parameter space that allows a late time de-Sitter attractor compatible with observations and a successful Vainsthein mechanism. We find that the requirement on having a successful Vainsthein mechanism is not compatible with the existence of cosmological solutions at early times.
\end{abstract}
\maketitle


\section{Introduction}
Einstein's theory of general relativity (GR) \cite{Einstein1916} has been the widely accepted theory of gravity with impeccable ability to match observations for over a century \cite{Will:2005va}. However, the discovery of the accelerated expansion of the universe \cite{Riess:1998cb,*Perlmutter:1998np} has lead to the construction of many modified theories of gravity which attempt to account for this observation in a more natural way than the addition of a cosmological constant to the Einstein field equations. See \cite{Clifton:2011jh,Joyce:2014kja,Koyama:2015vza} for reviews.

Dropping the notion of a massless spin-2 graviton is arguably the natural extension to GR. The effect of endowing the graviton with a non-zero mass was first considered in 1939 by Fierz and Pauli in a linear construction \cite{Fierz:1939ix}. In this theory, the mass term is built by requiring the absence of negative energy states (ghosts) and breaks the linearised diffeomorphism invariance, thus resulting in a massive spin-2 field theory propagating 5 degrees of freedom. Following this initial work, van Dam and Veltman \cite{V1}, and independently, Zakharov \cite{Zakharov:1970cc} showed that in the massless limit of the linear theory, GR is not recovered. The cause of this discrepancy is that the helicity-0 component of the graviton does not decouple from the trace of the stress tensor of the matter source in this limit. The resolution to this problem, offered by Vainshtein, is to extend the theory to include non-linear self interactions to allow a smooth GR limit \cite{Vainshtein}.

However, non-linear theories generically suffer from the Boulware-Deser ghost, the unwanted $6$th degree of freedom which breaks the linear tuning of Fierz and Pauli \cite{PhysRevD.6.3368}. The ghost-free potential was constructed after almost $40$ years, by de Rham, Gabadadze and Tolley (dRGT) in a decoupling limit \cite{PhysRevD.82.044020,*PhysRevLett.106.231101} and was subsequently shown to be ghost free to all orders \cite{Hassan:2011hr}. The theory is built out of a tensor $\sqrt{g^{-1}f}$ constructed from a physical metric $g_{\mu \nu}$ and a flat fiducial metric $f_{\mu \nu}$.
In the context of cosmology, massive gravity with dRGT potentials do not allow exact cosmological solutions without generating pathologies. The self-accelerating branch suffers from non-linear ghost instability \cite{DeFelice:2012mx,*DeFelice:2013awa}, while the normal branch does not allow expansion \cite{DAmico:2011eto} in the case of flat fiducial metric. Extending the flat fiducial metric to maximally symmetric space-times \cite{Hassan:2011tf}, the normal branch can support cosmology. However, in the case of de Sitter fiducial metric \cite{Langlois:2012hk}, these either suffer from a Higuchi ghost \cite{Higuchi:1986py} or does not have successful Vainshtein mechanism \cite{Fasiello:2012rw}, while for anti-de Sitter the cosmology is protected against acceleration \cite{MartinMoruno:2013gq}.

The situation is less severe for the bigravity theory, where the $f_{\mu\nu}$ metric is promoted to be dynamical \cite{Hassan:2011zd}. In the model in which matter is coupled only to the physical metric $g_{\mu \nu}$, although the self-accelerating branch cannot evade the conclusions of Refs.\cite{DeFelice:2012mx,*DeFelice:2013awa}, the normal branch can sustain cosmology. For a late time acceleration that is sourced by a massive graviton, the mass needs to be generically of the order of Hubble rate today. For this scenario Ref.\cite{Comelli:2012db} showed that the scalar perturbations in the radiation dominated era suffer from a gradient instability, ruling out a viable cosmology. \footnote{
However the rapid growth of perturbations may be screened by cosmological \cite{Aoki:2015xqa} or standard \cite{Mortsell:2015exa} Vainshtein mechanism.}
There have been various studies on the stability of this model \cite{Konnig:2014dna, Konnig:2014xva, Lagos:2014lca, Cusin:2014psa,Comelli:2014bqa} and two ways to circumvent this conclusion have been proposed. The first is to impose a hierarchy between scales, effectively decoupling the massive graviton from the matter sector, thus making the model indistinguishable from GR \cite{Akrami:2015qga}. The second way is the so-called ``low energy limit'' \cite{DeFelice:2013nba}, where the bare mass parameter is allowed to be large $m\gg H_0$, while the late time accelerated expansion can be achieved via a fine tuning of coupling constants.  The stability of this model was shown for a large portion of the parameter space in \cite{DeFelice:2014nja}, while some implications for primordial gravitational waves were studied in \cite{Fasiello:2015csa}. 

In this work, we focus on the low energy limit model in which matter is coupled to a single metric. Although the model has interesting gravitational wave phenomenology \cite{DeFelice:2013nba}, its implications for cosmology has not been explored in detail. The goal of the present work is to fill this gap and determine the late time implications of this model. 

The paper is organised as follows; in Section \ref{model} we review the bigravity theory and give the equations of motion. In Section \ref{BG} we discuss the background evolution, focussing on the late time de Sitter attractor. In Section \ref{perturbation} we study both linear and second order scalar perturbations, to obtain the analogue Poisson's equation and to identify the Vainshtein radius, respectively. In Section \ref{tunings}, we summarise all the conditions obtained to fix three model parameters and discuss potential observable signatures in this reduced parameter space. We conclude with a discussion in Section \ref{sec:discussion}.

\section{Bigravity theory with dRGT interactions}\label{model}
In this Section, we give a brief review of bigravity. The action for the theory is given by \cite{Hassan:2011zd} 
    \begin{align}\label{action}
    \begin{split}
    S=\frac{M_g^2}{2}\int d^4 x \sqrt{-g}R[g]+\frac{M_f^2}{2}\int d^4 x \sqrt{-f}R[f] 
 +m^2 M_g^2 \int d^4 x \sqrt{-g}\sum_{n=0}^{4}\alpha_n\mathcal{L}_n +\int d^4 x \sqrt{-g}\mathcal{L}_{\rm matter},
 \end{split}
    \end{align}
where $\alpha_n$ are dimensionless parameters, $M_g$ and $M_f$ are the corresponding Planck scales for the two metrics $g$ and $f$. In the above, $\mathcal{L}_n$ are the dRGT interaction terms given by
\begin{align}
\mathcal{L}_0(\mathcal{K}) &= 1\,,\nonumber\\
\mathcal{L}_1(\mathcal{K}) &= [\mathcal{K}]\,,\nonumber\\
\mathcal{L}_2(\mathcal{K}) &= \frac{1}{2!}\,([\mathcal{K}]^2-[\mathcal{K}^2])\,,\nonumber\\
\mathcal{L}_3(\mathcal{K}) &= \frac{1}{3!}\,([\mathcal{K}]^3-3[\mathcal{K}][\mathcal{K}^2]+2[\mathcal{K}^3])\,,\nonumber\\
\mathcal{L}_4(\mathcal{K}) &= \frac{1}{4!}\,([\mathcal{K}]^4-6[\mathcal{K}]^2[\mathcal{K}^2]+8[\mathcal{K}][\mathcal{K}^3]+3[\mathcal{K}^2]^2-6[\mathcal{K}^4])\,,
\end{align}
where square brackets denote trace operation and we defined the building block tensor as
\begin{align}
        \mathcal{K}^\mu_{\;\;\nu} \equiv \delta^\mu_\nu -\left(\sqrt{g^{-1}f}\right)^\mu_{\;\;\nu}.
\end{align}
We note that the square-root above is a tensor operation defined by the relation
\begin{align}
        \left(\sqrt{g^{-1}f}\right)^\mu_{\;\;\rho}\left(\sqrt{g^{-1}f}\right)^\rho_{\;\;\nu} = g^{\mu \rho}f_{\nu \rho}\,.
\end{align}
In this setup $g_{\mu \nu}$ corresponds to the physical metric, i.e. the metric that matter sector $\mathcal {L}_{\rm matter}$ couples to, while $f_{\mu \nu}$ is a dynamical background field.

We now vary the action (\ref{action}) with respect to $g_{\mu \nu}$ and $f_{\mu \nu}$,
to yield the equations of motion for the $g$ and $f$ metrics, respectively:
\begin{align}\label{eom}
        \mathcal{E}^{(g)}_{\mu\nu} &\equiv G_{\mu \nu}-\frac{1}{M_g^2}T_{\mu \nu}-m^2\sum_{n=0}^{4}\alpha_n\left(g_{\mu \nu}\mathcal{L}_n-2\frac{\delta \mathcal{L}_n}{\delta g^{\mu \nu}}\right)=0\,,\nonumber\\
        \mathcal{E}^{(f)}_{\mu\nu} &\equiv \mathcal{G}_{\mu \nu} +\frac{2 m^2\sqrt{-g} M_g^2}{\sqrt{-f} M_f^2}\sum_{n=0}^{4}\alpha_n \frac{\delta \mathcal{L}_n}{\delta f^{\mu \nu}}=0\,,
\end{align}
where $G_{\mu\nu}$ and $\mathcal{G}_{\mu\nu}$ are the Einstein tensors built out of the $g$ and $f$ metrics, respectively, and
\begin{equation}
T_{\mu\nu} \equiv -\frac{2}{\sqrt{-g}}\,\frac{\delta }{\delta g^{\mu\nu}}\,(\sqrt{-g} \mathcal{L}_{\rm matter})\,,
\end{equation}
is the energy-momentum tensor for the matter sector.  We show the explicit result for the variation of $\mathcal{L}_n$ with respect to $g^{\mu \nu}$ and $f^{\mu \nu}$ in appendix \ref{sec:eom}. Using (\ref{variation}), one can verify that (\ref{eom}) matches the expressions in Ref.\cite{Babichev:2013pfa}.

\section{Background Cosmology}
\label{BG}
In this section we study the background cosmology under the ansatz that both metrics take Friedmann Lema\^itre Robertson Walker (FLRW) form in the same coordinate system:
\begin{align}
    ds_g^2 &=-N(t)^2+a(t)^2\delta_{ij}dx^i dx^j \nonumber\\
    ds^2_f &=-n(t)^2+\alpha(t)^2\delta_{ij}dx^i dx^j\,,
    \label{eq:metricbackground}
\end{align}
where $n(t)$, $N(t)$ denote the lapse functions, while $a(t)$, $\alpha(t)$ represent the scale factors for the $g$, $f$ metrics, respectively. For the matter content we consider a perfect fluid described by the energy-momentum tensor:
\begin{equation}
    T_{\mu \nu}=\rho u_{\mu} u_{\nu}+P(g_{\mu \nu}+u_{\mu}u_{\nu}),
    \label{eq:enmommatter}
\end{equation}
where $u_{\mu}$ is the 4-velocity of the fluid and satisfies the condition $u_{\mu}u^{\mu}=-1$, $P$ the pressure and $\rho$ the energy density. In accordance with the homogeneous and isotropic metric ansatze, the background values for the pressure and energy density are functions of time only, while $u_{\mu} = -\delta_\mu^0 N$. In what follows, we will restrict our discussion to a matter sector consisting only of a pressureless non-relativistic fluid with $P=0$. 

For the background metrics \eqref{eq:metricbackground} and a perfect fluid \eqref{eq:enmommatter}, the equations of motion \eqref{eom} reduce to four independent equations \cite{DeFelice:2014nja}
\begin{align}\label{friedmannG}
   3H^2&=m^2 \rho_{m,g}+\frac{\rho}{M_g^2} 
   \,,\\
   \label{friedmannF}
    3H_f^2 & =\frac{m^2}{\kappa} \rho_{m,f}
    \,,\\
    \label{accelg}
\frac{2\,\dot{H}}{N}& =m^2\xi\,J(\tilde{c}-1)-\frac{\rho}{M_g^2}
\,,\\
\label{accelf}
\frac{2\,\dot{H_f}}{n}& =-\frac{m^2}{\kappa \xi^3 \tilde{c}}J(\tilde{c}-1)
\,,\\
\label{matter}
\frac{\dot{\rho}}{N} + 3H \rho &=0,
\end{align}
where a dot represents a time derivative and we defined the following functions in accordance with the notation of Ref.\cite{DeFelice:2014nja}:
\begin{align}
\rho_{m,g}(\xi)&\equiv U(\xi)-\frac{\xi}{4}\partial_{\xi}U(\xi)\,,\nonumber\\
\rho_{m,f}(\xi)&\equiv \frac{1}{4\xi^3}\partial_{\xi}U(\xi)\,,\nonumber\\
J(\xi) &\equiv \frac{1}{3} \partial_{\xi}\left(U(\xi)-\frac{\xi}{4}\partial_{\xi}U(\xi)\right)\,,
\end{align}
with $U(\xi)\equiv-\alpha_0+4\alpha_1(\xi-1)-6(\xi-1)^2+4\alpha_3(\xi-1)^3-\alpha_4(\xi-1)^4$ and we also have
\begin{equation}
\xi\equiv \frac{\alpha}{a}\,,\qquad
\tilde{c}\equiv \frac{n\,a}{N\,\alpha}\,,\qquad
\kappa \equiv M_f^2/M_g^2 \,.
\end{equation}

The contracted Bianchi identity for individual metrics yields an effective constraint. For instance, differentiating \eqref{friedmannG} then combining it with \eqref{accelg} and \eqref{matter}, we obtain
\begin{equation}
    J(H-\xi H_f)=0\,.
\end{equation}
This equation branches out into two solutions, where the self-accelerating branch $J=0$ is known to lead to non-linear ghost instabilities \cite{DeFelice:2012mx,*DeFelice:2013awa}. Instead, we choose the normal branch with $H=\xi H_f$. This solution links the evolution of the $f$ metric to the $g$ metric one and the consistency of the two Friedmann equations give an algebraic relation between $\xi$ and the matter density $\rho$
\begin{equation}\label{xi}
    \hat{\rho}_m(\xi)=-\frac{\rho}{m^2 M_g^2}\,,
\end{equation}
where we defined the combination
\begin{equation}
\hat{\rho}_m(\xi)\equiv \rho_{m,g}-\frac{\xi^2}{\kappa}\rho_{m,f}\,.
\end{equation}
In order to avoid the early time gradient instability in this branch \cite{Comelli:2014bqa}, we will adopt the low energy limit defined by
\begin{equation}
\rho \ll m^2 M_g^2\,,
\label{eq:lowenergy}
\end{equation}
which allows us to push the instability beyond the reach of the effective field theory \cite{DeFelice:2013nba,DeFelice:2014nja}. As time evolves, the density $\rho$ redshifts as $a^{-3}$, and the solution for $\xi$ converges to a constant value $\xi_c$ defined by
\begin{equation}
    \hat{\rho}_m(\xi_c)=0\,.
    \label{desitter}
\end{equation}
To describe the evolution close to this late time attractor, we linearise Eq.(\ref{xi}) around $\xi=\xi_c$ to relate the departure from this point to the matter density:
\begin{align}
    m^2\left[\frac{3(1+\kappa \xi_c^2)J_c}{\kappa \xi_c^2} -
\frac{2\,\Lambda}{m^2 \xi_c}
\right](\xi-\xi_c) \sim \frac{\rho}{M_g^2}\,,
\label{eq:linearapprox}
\end{align}
where $\Lambda\equiv m^2\rho_{m,g}(\xi_c)$
and the subscript $c$ corresponds to the values of functions evaluated at the de-Sitter attractor. Following Ref.\cite{DeFelice:2014nja}, we now assume $\vert\Lambda/(m^2J_c)\vert \ll 1$. Using these results, we find that the Friedmann equation (\ref{friedmannG}) can be approximated as
\begin{align}
    3H^2 \simeq \frac{\rho}{\tilde{M}_g^2}+\Lambda\,,
    \label{eq:approxfriedmann}
\end{align}
where the effective Planck scale is $\tilde{M}_g^2\equiv (1+\kappa \xi_c^2)M_g^2$, and we now identify $\Lambda$ as the effective cosmological constant \footnote{Notice that $\Lambda$ has a contribution from $\alpha_0$, which is simply a bare cosmological constant. In Sec.\ref{tunings}, we will set $\alpha_0=0$  such that the accelerated expansion is solely due to the two-metric coupling.}.

\section{Cosmological perturbations}\label{perturbation}
In this section we consider perturbations around the low energy background model and determine the effect of the two-metric interaction on the linear growth of structure. We outline the method taken to study the perturbations in theory, the process to isolate the scalar mode in the Poisson's equation and the extension to non-linear order. We only consider scalar perturbations in this work as they are the only relevant ones for the large scale structure. 
\subsection{Linear perturbations}
To begin, we perturb both metrics around the backgrounds \eqref{eq:metricbackground} in the Newtonian gauge for the $g$ metric
\begin{align}
ds_g^2 &=-(1+2\,\phi)dt^2
+a^2\left(\delta_{ij}+2\,\delta_{ij}\psi\right)dx^i dx^j\,,\nonumber\\
ds_f^2 &=-n^2(1+2\,\phi_f)dt^2+2\,n\,a\,\partial_i b \,dt \,dx^i  +\alpha^2\left[\delta_{ij}+2\,\delta_{ij}\psi_f+\left(\partial_i \partial_j-\frac{\delta_{ij}}{3}\nabla^2\right)S\right]dx^i dx^j\,,
\end{align}
where $(\phi,\psi,b,S,\phi_f,\psi_f)$ are the perturbation variables and we fixed the time coordinate such that $N=1$. The perturbations in the matter sector are introduced via $\rho=\rho(t)+\delta \rho$ and $u^\mu= (1-\phi, \partial^iv)$, giving
\begin{align}
T_{00}&= \rho\, \left(1+2\,\phi+\frac{\delta\rho}{\rho}\right)\,,\nonumber\\
T_{0i}&= -a^2\,\rho \,\partial_iv\,,\nonumber\\
T_{ij}&= 0\,.
\end{align}
With these decompositions and using the quasi-static approximation \cite{Sawicki:2015zya} we can derive an analogue of  Poisson's equation for the potential $\phi$ \cite{DeFelice:2014nja}
\begin{align}\label{Poisson}
     \phi=-\frac{\delta \rho}{2 \tilde{M}_g (k^2/a^2)}\left[\frac{6W+(3+4\kappa \xi_c^2)(k^2/a^2)}{6W+3(k^2/a^2)}\right]\,,
\end{align}
where $k$ corresponds the momentum of the mode in the plane-wave expansion. The derivation is summarised in appendix~\ref{sec:poisson}.
The contribution from the two-metric interaction is encoded in the function $W$ defined by
\begin{align}
    W\equiv\frac{m^2(1+\kappa \xi_c^2)J}{2\kappa \xi_c}-H^2\,.
    \label{eq:higuchi}
\end{align}
This quantity also plays a major role in the perturbative stability conditions, with $W>0$ corresponding to the bigravity generalisation of the Higuchi bound \cite{DeFelice:2014nja}.

The form of the Poisson's equation is similar to the one in the presence of a scalar field source. The traceless part of the $g$ equations of motion given by  \eqref{Gtrless}, 
\begin{equation}
\label{traceless}
J_cm^2 \xi_c a^2 S+2(\phi+\psi) =0,
\end{equation}
reveals that the perturbation $S$ acts as a source for anisotropic stress. 

\subsection{Vainshtein radius}
We now move on to the study of second order perturbations and an identify the scale at which the perturbative expansion breaks down. This will allow us to determine the Vainshtein radius where the scalar graviton decouples from the matter sector and the evolution closely follows GR. 

In order to do this, a few approximations are in order. In addition to restricting the study to the de Sitter attractor, we consider scales where the expansion can be neglected. We also focus on small scales and assume $\nabla^2 \gg m^2$. Keeping metric perturbations up to second order in the equations of motion, we find that unlike the $g$-metric equations, the non-linear $f$-metric equations do not exhibit the enhancement $\nabla^2/m^2$ with respect to the linear part. This allows us to solve the linear $f$-metric equations and substitute the solutions for $(\psi_f,\phi_f,\phi)$ into the non-linear components of the g-metric equations. The solutions are;
\begin{align}
\label{phifpsifphi}
    \phi_f &=\frac{J_cm^2S}{4 \kappa \xi_c}\,,\nonumber\\
    \psi_f &=\frac{1}{6}\nabla^2 S+\frac{J_cm^2 S}{4\kappa \xi_c}\,,\nonumber\\
    \phi &=\frac{3\,J_c m^2S}{ 4\kappa \xi_c}-2\psi\,,
\end{align}
Upon substitution of equations \eqref{phifpsifphi} into the g-metric equations we obtain
\begin{align}\label{G001}
   \mathcal{E}^{(g)}_{00}&= \frac{m^2 \xi_c^3}{16(\xi_c-1)}\left[(\nabla^2 S)^2-\left(\partial_{i}\partial_{j}S \partial^{i}\partial^{j}S\right)\right]
    +\frac{1}{2}J_cm^2\xi_c\nabla^2S+2\nabla^2 \psi+\frac{\delta \rho(t)}{M_g^2}
\,,\\
    \label{Gtr1}
    \delta^{ij}\mathcal{E}^{(g)}_{ij}&= \frac{m^2 \xi_c^3}{16(\xi_c-1)}\left[(\nabla^2 S)^2-\left(\partial_{i}\partial_{j}S \partial^{i}\partial^{j}S\right)\right]+
   J_c m^2 \xi_c\nabla^2 S-2\nabla^2 \psi +\frac{3J_c^2 \nabla^2 S}{4\kappa \xi_c }\,,
\end{align}
where we kept only the second order terms that are enhanced in the limit $\nabla^2/J_c m^2 \gg 1$. Using Eq.~(\ref{Gtr1}) to replace $\nabla^2\psi$, Eq.~(\ref{G001}) reduces to
\begin{align}
    \frac{m^2}{8\xi_c}\left\{\frac{\xi_c^4}{(\xi_c-1)(1+\kappa \xi_c^2)}\left[(\nabla^2 S)^2-\left(\partial_{i}\partial_{j}S\right)^2\right]\right\}
    +\frac{m^2}{8\xi_c}\left(\frac{12 J_c}{\kappa}\nabla^2 S\right)+\frac{\delta \rho}{\tilde{M}_g^2}=0\,,
    \label{eq:nonlinearS}
\end{align}
where the non-linear term has the expected galileon-like structure. The scale at which the non-linear terms become important depends on the normalisation of the field. We define
\begin{equation}
\tilde{S} = -\frac{m^2 J_c\xi_c}{2}\,S\,,
\end{equation}
such that the linear traceless equation of motion \eqref{traceless} reduces to $\tilde{S} = \psi+\phi$. With this normalisation, the non-linear equation \eqref{eq:nonlinearS} becomes
\begin{align}
    \nabla^2 \tilde{S}
    -\frac{C}{6}\,\left[(\nabla^2 \tilde{S})^2-\left(\partial_{i}\partial_{j}\tilde{S}\right)^2\right]=\frac{\kappa\,\xi_c^2\delta \rho}{3\,\tilde{M}_g^2}\,,
\end{align}
where 
\begin{equation}
C\equiv\frac{\kappa \xi_c^3}{J_c^2 m^2(\xi_c-1)(1+\kappa \xi_c^2)}\,.
\label{eq:Cfactor}
\end{equation}
Thus the non-linear term dominates for $ C \nabla^2 \tilde{S} \sim \mathcal{O}(1)$, revealing the order of the Vainshtein radius as
\begin{equation}
    R_v \sim (C\, r_g)^{\frac{1}{3}}\,,
    \label{eq:Vainshtein}
\end{equation}
where $r_g$ corresponds to the Schwarzschild radius of a spherical body.
This result is consistent with the similar calculation in Ref.\cite{DeFelice:2013nba}.

\section{Fixing Model Parameters}\label{tunings}
We are now ready to fix the model parameters without introducing the bare cosmological constant term, which is equivalent to setting $\alpha_0=0$. On the other hand, we keep $\alpha_1$ non-zero. This means that the theory does not admit the Minkowski solution, which is not a problem as we are interested in cosmological solutions.

We start by trading $\alpha_3$ for $\xi_c$ using the definition of the de-Sitter fixed point (\ref{desitter}) and obtain $\alpha_3= \alpha_3(\xi_c,\alpha_1,\alpha_4,\kappa)$ as 
\begin{align}
\alpha_3=\frac{
	3(\xi_c-1)\left(\xi_c -2\,-\frac{1}{\kappa\,\xi_c}\right)
	+\left(4 +\frac{1}{\kappa\,\xi_c} -3\,\xi_c\right)\,\alpha_1
	-(\xi_c-1)^3\left(\frac{1}{\kappa\,\xi_c}+1\right)\alpha_4}{(\xi_c-1)^2 \left[\xi_c-4-\frac{3}{\xi_c \kappa}\right]}
\end{align}
We then fix $\alpha_4$ matching the effective cosmological constant $\Lambda$ in the approximate Friedmann equation \eqref{eq:approxfriedmann} to the observed value, which is equivalent to solving,
\begin{equation}
\rho_{m,g}=\left(\frac{H_0}{m}\right)^2\,.
\end{equation}
The solution for $\alpha_4=\alpha_4(\xi_c,\alpha_1,\kappa)$ is, 
\begin{align}
\alpha_4 = \frac{6}{(\xi_c-1)^2}-\frac{8\,\alpha_1}{(\xi_c-1)^3}+\left(\frac{H_0}{m}\right)^2\frac{3-\kappa\xi_c(\xi_c-4)}{(\xi_c-1)^4}.
\end{align}
Finally, the last parameter is fixed by requiring a sensible Vainshtein radius which ensures that the effects of modified gravity are hidden below a certain distance scale to recover GR on solar system and galactic scales. 
The workings of the Vainshtein mechanism are that derivative self interactions of the scalar are enhanced around a matter source such as a star, so that the effect of the fifth force is screened below the Vainshtein radius. The relation we will consider is,
\begin{equation}
R_v^3=r_c^2r_g\,.
\end{equation}
We introduce the parametrisation  
\begin{equation}
r_c=b\,H_0^{-1}\,,
\label{eq:introduceb}
\end{equation}
where $b \sim \mathcal{O}(0.1-1)$ \cite{Brax:2011sv, Sakstein:2017bws} and allows us to tune the size of the Vainshtein radius. The equation which relates the model parameters to the Vainshtein radius is \eqref{eq:Vainshtein}
\begin{equation}\label{Vainshtein}
C=\frac{b^2}{H_0^2}.
\end{equation}
From this relation we then fix $\alpha_1$,
%
\begin{equation}
 \alpha_1 = \frac{\xi_c-1}{2} \pm \frac{\sqrt{\kappa \xi_c ( \xi_c-1)}}{2\,b\,\sqrt{1+\kappa\xi_c^2}} \frac{H_0}{m} + \mathcal{O}\left(\frac{H_0}{m}\right)^2\,,
\end{equation}
where we made use of the fact that $H_0 \ll m$. Noting that the solution with the $+$ sign leads to a negative $W$ (defined in Eq.\eqref{eq:higuchi}), we will choose the $-$ sign solution in the following. Moreover, the solution only exists for $\xi_c>1$. 

After this procedure, we have reduced the number of free parameters down to two: $\xi_c$ and $\kappa$. 
We now check whether there are any inconsistencies in the background equations of motion. Expanding the left hand side of Eq.\eqref{xi} around the attractor, we have 
\begin{align}
\frac{d \hat{\rho}_m}{d\xi}\biggr\rvert_{\xi=\xi_c}(\xi-\xi_c)+\frac{d^2\hat{\rho}_m}{d\xi^2}\biggr\rvert_{\xi=\xi_c}(\xi-\xi_c)^2 + \mathcal{O}(\xi-\xi_c)^3=-\frac{\rho}{m^2 M_g^2}\,.
\label{eq:quadratic}
\end{align}
In the limit $H_0\ll m$, the coefficients of the linear and quadratic terms are,
\begin{align}
\frac{d \hat{\rho}_m}{d\xi} \Bigg\vert_{\xi=\xi_c} &= \frac{3\,\sqrt{1+\kappa\,\xi_c^2}}{b\,\sqrt{\kappa\,\xi_c(\xi_c-1)}}\,\frac{H_0}{m}+\mathcal{O}\left(\frac{H_0}{m}\right)^2\,,\nonumber\\
\frac{d^2 \hat{\rho}_m}{d\xi^2} \Bigg\vert_{\xi=\xi_c} &= 
\frac{3\,(1+\kappa\,\xi_c^2)}{\kappa\,\xi_c(\xi_c-1)}
+\mathcal{O}\left(\frac{H_0}{m}\right)\,.
\end{align}
In Section \ref{BG}, we used the linear term to obtain the approximate Friedmann equation \eqref{eq:approxfriedmann}.  However, we see that the first derivative is suppressed by $H_0/m$, while the second derivative term is manifestly positive. As a result, when the quadratic term dominates, there is no real solution to this equation. This observation allows us to determine the parameter range which grants a physical evolution.  The linear term is dominant if 
\begin{equation}
 \vert\xi-\xi_c \vert \lesssim \frac{1}{b}\,\frac{H_0}{m}\,,
 \label{eq:lineardom}
\end{equation}
in which case, the solution to \eqref{eq:quadratic} behaves as
\begin{equation}
    \vert\xi-\xi_c\vert\sim b\left(\frac{H_0}{m}\right)^{-1}\frac{\rho}{m^2 M_g^2}\,.
\end{equation}
Using the condition \eqref{eq:lineardom}, the above relation yields  an upper bound on $b$;
\begin{align}
    b  <  \frac{H_0M_g}{\sqrt{\rho}}\,.
\end{align}
Since the matter density today is of order of $H_0^2 M_g^2$, we use $\rho \sim H_0^2M_g^2/a^3$, giving 
\begin{align}
    b < a^{3/2}\,.
\end{align}
Therefore, the solution exists for 
\begin{equation}
a>a_{\rm in} = b^{3/2}\,.
\label{b}
\end{equation}
Turning this relation around, given a parameter $b$, the cosmological description can go as far back as $a_{\rm in}$, before which no physical evolution exists. Although we set $\alpha_0 =0$ in order not to introduce a bare cosmological constant, we can check that this conclusion holds even if $\alpha_0 \neq 0$.

\begin{figure}
    \centering
    \includegraphics[width=0.64\textwidth]{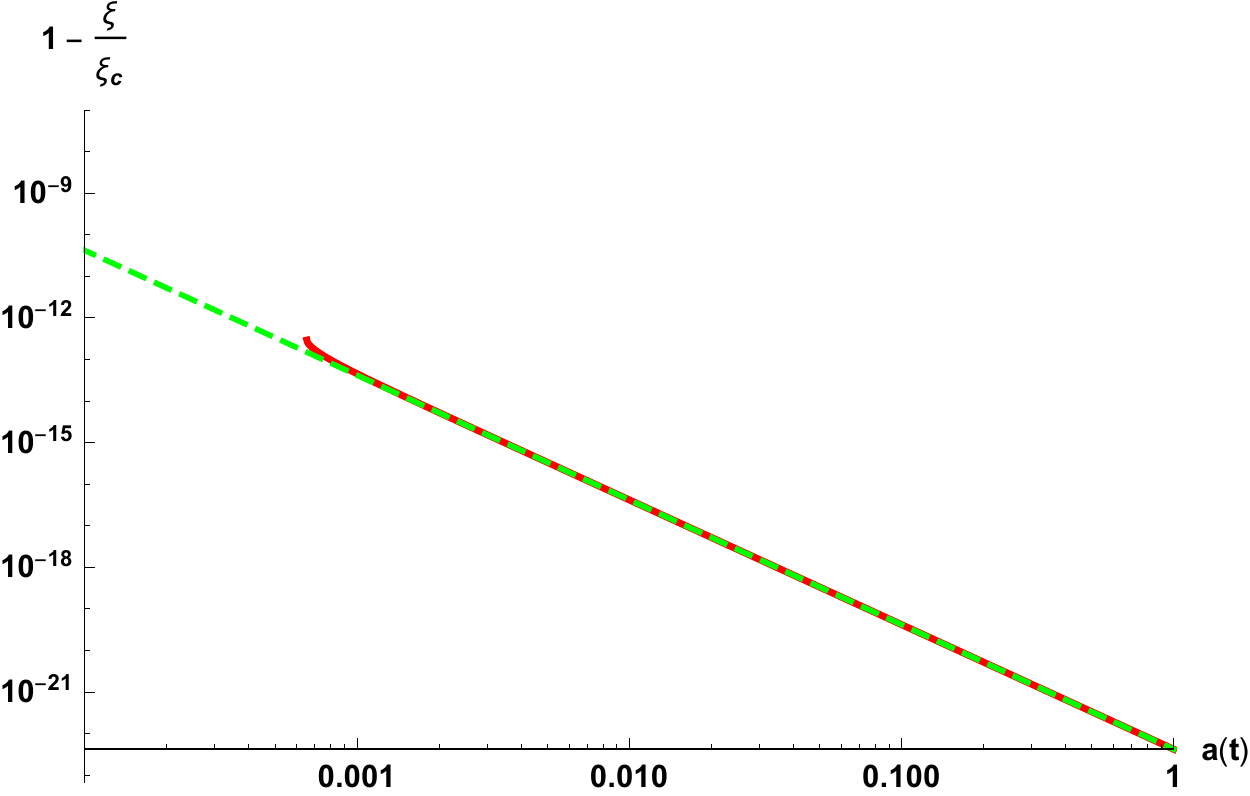}
    \caption{The evolution of $1-\xi/\xi_c$ with the scale factor. The dashed green line shows the linear approximation \eqref{eq:linearapprox} while the solid red line corresponds to the numerical solution obtained by solving the exact equation \eqref{xi}. The Vainshtein radius tuning parameter, defined in Eq.\eqref{eq:introduceb}, is $b=10^{-9/2}$, while the other parameters are set to $\kappa=1$, $\xi_c=8$.}
    \label{fig:xi-evol}
\end{figure}

Suppose we wish to describe the evolution of the scale factor up from the last scattering surface onward. Therefore, we set $a_{\rm in}=a_{\rm CMB}=10^{-3}$. In order to have the low energy limit \eqref{eq:lowenergy} be valid at the time of CMB, the minimum mass parameter allowed is $m=H_{\rm CMB}$, where $H_{\rm CMB}$ is the Hubble parameter at the last scattering surface. From Eq.~(\ref{b}), this initial value of the scale factor corresponds to a value of $b=10^{-9/2}$. In order to check this estimate, we compared the exact numerical solution of Eq.\eqref{xi} to the linear approximation. The comparison is summarised in figure \ref{fig:xi-evol}. The exact solution only appears around $a\sim 10^{-3}$, after which the value of $\xi$ becomes closer to the de Sitter attractor value $\xi_c$.

We close this Section with a discussion of the effect on the large scale structure. From Eq.~(\ref{Poisson}) we can determine the consequences of the several tunings: in order to have an observable effect, the quantity $W$ has to be comparable to the $k^2$ contribution. The function $W$ can be interpreted as the effective mass of the gravity perturbations, and behaves as $W \sim m^2 J$. It encodes the information about the scale at which the modifications to gravity appear. Using the approximate expression,
\begin{align}
    W \sim m^2 \left[\left(\frac{1}{b}\frac{H_0}{m}\right)+\mathcal{O}\left(\frac{H_0}{m}\right)^2\right]\,,
\end{align}
we can estimate its value, using $m\sim H_{CMB} \sim 10^{9/2} H_0$,
 \begin{equation}
     W \sim \frac{1}{\left(10^{-3/2}{\rm Mpc}\right)^2} \left(\frac{10^3 {\rm Mpc}}{H_0^{-1}}\right)^{2} \,,
 \end{equation}
which implies that the effect of the two-metric interaction appears only at scales smaller than $\sim 0.1 {\rm Mpc}$ where linear perturbation theory is no longer applicable. 

\section{Discussion and Conclusions}
\label{sec:discussion}
We have presented an analysis of the linear and non-linear perturbations in bigravity where non-derivative two metric coupling is introduced as in \cite{Hassan:2011zd} so as not to generate the Boulware-Deser ghost. We considered a perfect fluid with equation of state $P=0$ coupled to the g-metric and studied metric perturbations around FLRW, whilst adopting the healthy branch of solution with $H=\xi H_f$. Poisson's equation was derived at the linear level in perturbations and we identified the modification to the Poisson's equation due to the extra degrees of freedom present from the massive graviton. Furthermore, we studied peturbations going beyond linear order and identified the Vainshtein radius, below which the derivative self-interactions of the scalar screen the effect of the fifth force and conspire to reproduce GR on local scales.

We then looked at the effect of fixing three of the model parameters has to the background cosmology of the theory using the following requirements: ensuring the existence of the late-time de-Sitter attractor, matching the effective cosmological constant in the Friedmann equation to the value we observe and proposing we have a Vainshtein radius. Bigravity in the low energy limit can admit a sensible cosmological solution, but this comes with a cost of lowering the Vainsthein radius. With a value of $b=10^{-9/2}$, for which the Vainshtein radius is given by $R_v^3=10^{-9} H_0^{-2} r_g$, we are able to describe the evolution of the scale factor up until the last scattering surface at $a_{\rm in}=10^{-3}$. However, to satisfy the observational constraints, we need to impose $b={\cal O}(0.1-1)$, which results in a very short window of the viable cosmological evolution.

The main conclusion of this paper is that the stable bigravity model that is distinguishable from GR does not provide a reasonable description for the late time acceleration of the universe. With this result, we have established that none of the exact cosmological solutions to dRGT massive gravity/bigravity theory, where matter couples to a single metric, admit a viable and testable dark energy model. There are several potential directions from here. Most conservative option is to explore extensions which preserve the local Lorentz invariance of dRGT, while avoiding new dynamical degrees of freedom.

One such extension is to allow matter to couple to a composite metric which avoids the generation of Boulware-Deser ghost within the range of the effective field theory \cite{deRham:2014naa}. In the normal branch of cosmological evolution \cite{Enander:2014xga}, the vector modes suffer from a gradient instability in the radiation era \cite{Comelli:2015pua}, while the scalar mode becomes a ghost in the late universe \cite{Gumrukcuoglu:2015nua}. The self-accelerating branch, which would be problematic in the non-composite theory, becomes detuned by the effect of matter and allows a stable evolution that undergoes a bounce \cite{Gumrukcuoglu:2015nua,Lagos:2015sya}. For the composite coupled theories, if the double coupling extends to the Standard Model sector, this implies that the light cone corresponding to the observed gravitational waves from binary neutron star merger \cite{Monitor:2017mdv} is different than the one for photons \cite{Akrami:2018yjz}.  

Another extension that persists is the generalised massive gravity, where the translation symmetry in the St\"uckelberg scalar field space is broken, while the Lorentz invariance remains intact \cite{deRham:2014gla}. In this construction, the number of degrees of freedom are the same as in dRGT although the parameters are promoted to functions of the scalar fields. This theory admits self-accelerating open universe solutions and their stability was shown in the decoupling limit. However, its phenomenology remains largely unexplored. 
\newline
\newline 
\textbf{Note Added}: Shortly after this work was sumbitted to the arXiv, Ref.\cite{Luben:2018ekw} appeared on the arxiv claiming our conclusions on the viability of bigravity cosmology cannot hold in general. As we clearly stated in the abstract and the rest of the paper, we are considering the low energy limit of the theory and our aim was never to find a general conclusion, nor did we make this claim. Both of the counter-examples that were highlighted in Ref.\cite{Luben:2018ekw} (i.e. hierarchy between the Planck scales, and screening of gradient instabilities) are already mentioned in the Introduction section.

\acknowledgments
We thank Matteo Fasiello for useful discussions. The work of AEG and KK has received funding from the European Research Council
(ERC) under the European Union’s Horizon 2020 research and innovation programme (grant agreement No. 646702 ”CosTesGrav”). KK is supported by the UK STFC grant ST/N000668/1. 

\appendix
\section{Derivation of equations of motion}
\label{sec:eom}
In this appendix, we compute the variation of the interaction term with respect to $g^{\mu \nu}$ and $f^{\mu \nu}$. 
We define 
\begin{align}
X^{\alpha}_{\;\;\beta} =\left(\sqrt{g^{-1}f}\right)^\alpha_{\;\;\beta}, \quad 
X^\alpha_{\;\;\beta}X^\beta_{\;\;\gamma} = g^{\alpha\beta}f_{\beta\gamma}\,.
\end{align}
Using this definition, we can vary the trace of various powers of this tensor:
\begin{equation}
\delta [X^n] = \frac{n}{2}\,(X^n)^\alpha_{\;\;\mu}\left(g_{\alpha\nu}\delta g^{\mu\nu} - f_{\alpha\nu}\delta f^{\mu\nu}\right)\,,
\end{equation}
which is valid for any power $n\ge 1$. In the above we made use of $\delta f_{\mu\nu} = - f_{\mu\alpha}f_{\nu\beta}\delta f^{\alpha\beta}$.
The variation of the interaction terms can be written in the following form:
\begin{align}
\label{variation}
\delta \mathcal{L}_1 &= -\frac{1}{2}\, X^\alpha_{\;\;\mu} \left(g_{\alpha\nu}\delta g^{\mu\nu} - f_{\alpha\nu}\delta f^{\mu\nu}\right)\,,\nonumber\\
\delta \mathcal{L}_2 &= \left[\left(-\frac{3}{2}+\frac{1}{2}[X]\right)X - \frac{1}{2} X^2 \right]^\alpha_{\;\;\mu}
\left(g_{\alpha\nu}\delta g^{\mu\nu} - f_{\alpha\nu}\delta f^{\mu\nu}\right)\,,\nonumber\\
\delta \mathcal{L}_3 &= \left[\left(-\frac{3}{2}+[X]-\frac{1}{4}[X]^2+\frac{1}{4}[X]^2 \right)X + \left(-1+\frac{1}{2}[X]\right) X^2 -\frac{1}{2}X^3\right]^\alpha_{\;\;\mu}
\left(g_{\alpha\nu}\delta g^{\mu\nu} - f_{\alpha\nu}\delta f^{\mu\nu}\right)\,,\nonumber\\
\delta \mathcal{L}_4 &= \left[\left(-\frac{1}{2}+\frac{1}{2}[X]-\frac{1}{4}[X]^2+\frac{1}{12}[X]^3+\frac{1}{4}[X^2] -\frac{1}{4}[X][X^2]+\frac{1}{6}[X^3] \right)X \right.\nonumber\\
& \left. \qquad+ \left(-\frac{1}{2} +\frac{1}{2}[X]-\frac{1}{4}[X]^2+\frac{1}{4}[X^2]\right) X^2 +\left(-\frac{1}{2}+\frac{1}{2}[X]\right)X^3-\frac{1}{2}X^4\right]^\alpha_{\;\;\mu}
\left(g_{\alpha\nu}\delta g^{\mu\nu} - f_{\alpha\nu}\delta f^{\mu\nu}\right)\,.
\end{align}

\section{Derivation of Linear Poisson's equation}
\label{sec:poisson}
To obtain the Poisson's equation we first substitute the perturbed metrics into the equations of motion, whilst in the process setting all time derivatives of the fields to zero in accordance with the quasi-static approximation and evaluating everything at the late time attractor. The equations take the following form,  up to non-zero factors
\begin{align}
\mathcal{E}^{(g)}_{00} &\rightarrow\frac{\delta \rho}{M_g^2}-\frac{2k^2\psi}{a^2}-3J_cm^2\xi_c(\psi-\psi_f),
\label{G00} \\
\delta^{ij} \mathcal{E}^{(g)}_{ij} & \rightarrow\left(-3J_cm^2 \xi_c-2(k^2/a^2)\right)\phi
    +3J_cm^2 \xi_c\phi_f-6J_cm^2\xi_c\psi-2(k^2/a^2)\psi+6J_cm^2\xi_c\psi_f,
\label{Gij}  \\
    \mathcal{E}^{(g)}_{ii}-(1/3)\delta^{ij}\mathcal{E}^{(g)}_{ij}&\rightarrow J_cm^2 \xi_c a^2 S+2(\phi+\psi),
\label{Gtrless}   \\
\mathcal{E}^{(f)}_{00} &\rightarrow -k^2\kappa\xi_c(k^2S+6\psi_f)+a^2(9J_cm^2\psi-9J_cm^2\psi_f),
\label{F00} \\
    \delta^{ij}\mathcal{E}^{(f)}_{ij}&\rightarrow -3a^2(3J_c m^2 \phi -3J_cm^2 \phi_f+6J_cm^2 \psi- 6J_c m^2 \psi_f)
    +k^2 \kappa \xi_c (k^2 S+6(\phi_f+\psi_f)),
\label{Ftrace} \\
\mathcal{E}^{(f)}_{ii}-(1/3)\delta^{ij}\mathcal{E}^{(f)}_{ij}&\rightarrow -3J_c m^2a^2S+\kappa \xi_c(k^2 S+6(\phi_f+\psi_f)).
 \label{Ftrless} 
 \end{align}
 We then solve equations (\ref{Gtrless}, \ref{F00}, \ref{Ftrace}, \ref{Ftrless}) for the variables $(S,\phi_f,\psi_f,\psi)$. The explicit solutions are as follows;
 \begin{align}
     \psi & =\frac{1}{2} \left(-J_c m^2 \xi_c a^2 S-2 \phi \right),
\label{psi} \\
     \psi_f &=-\frac{\xi_c S \left(9 J_c^2 m^4 a^4+2 \kappa  k^4\right)+18 J_c m^2 a^2 \phi }{6 \left(3 J_c m^2 a^2+2
   \kappa  k^2 \xi_c\right)},
 \label{psif} \\
\phi_f &=
\frac{J_c m^2 a^2 \left[\left(3 J_c m^2 a^2 (1+\kappa  \xi_c^2)+\kappa  \xi_ck^2 \right)S  + 6 \kappa
   \xi_c \phi\right] }{2 \kappa  \xi_c \left(3J_c m^2 a^2+2 \kappa  k^2 \xi_c\right)},
\label{phif} \\
     S &=-\frac{4 \kappa ^2 k^2 \xi_c^2 \phi }{J_c m^2 a^2 \left(3 J_c m^2 a^2 \left(\kappa  \xi_c^2+1\right)+\kappa 
   k^2 \xi_c \left(4 \kappa \xi_c^2+3\right)\right)}.
\label{S}
 \end{align}
 Substituting solutions(\ref{psi}-\ref{S}) into (\ref{G00}) and performing the re-definitions 
 \begin{align}
 M_g=\frac{\tilde{M}_g}{\sqrt{1+\kappa \xi_c^2}}, \quad  J_c=\frac{2\kappa \xi_c W}{m^2(1+\kappa \xi_c^2)},
 \end{align}
yields equation (\ref{Poisson}). Note that the trace part of the g-metric equations (\ref{Gij}) is automatically satisfied.
\bibliography{ref1}

\end{document}